\documentclass[useAMS,usenatbib]{mn2e}

\usepackage{graphicx}
\usepackage{amssymb}
\makeatletter
\usepackage{times}
\usepackage{amsmath}

\usepackage{mathptmx}
\usepackage[T1]{fontenc}
\usepackage{verbatim}
\usepackage{framed}
\usepackage{color}
\definecolor{shadecolor}{rgb}{1,1,0.78}
\definecolor{shadecolor}{rgb}{1,1,0}

\newcommand{\Gyr}{{\,\rm Gyr}}

\newcommand{\pc}{\,\mathrm{pc}}

\newcommand{\Hz}{{\,\rm Hz}}

\newcommand{\Mbh}{M_{\bullet}}
\newcommand{\Ro}{R_{\odot}}
\newcommand{\Mo}{M_{\odot}}

\newcommand{\Ms}{M_{\star}}

\newcommand{\GW}{\rm GW}
\newcommand{\p}{p}
\newcommand{\M}{M}
\newcommand{\h}{h}
\newcommand{\cc}{c}
\newcommand{\s}{s}
\newcommand{\jj}{J}
\newcommand{\ii}{i}
\newcommand{\ttt}{t}
\newcommand{\MS}{\rm MS}

\newcommand{\apj}{ApJ}
\newcommand{\apjl}{ApJ}
\newcommand{\prd}{Phys. Rev. D}
\newcommand{\mnras}{MNRAS}

\newcommand{\aj}{AJ}

\title[Gravitational wave background]{The gravitational wave background from star-massive black hole fly-bys}

\author[Toonen, Hopman and Freitag]
{
Silvia Toonen$^{1,2}$ \thanks{e-mail: s.toonen@astro.ru.nl},
Clovis Hopman$^1$ and Marc Freitag$^3$\\
$^1$Leiden Observatory, Leiden University,  P.O. Box 9513, 2300 RA
Leiden, The Netherlands\\
$^2$Department of Astrophysics, IMAPP, Radboud University Nijmegen, P.O. Box 9010, 6500 GL Nijmegen, The Netherlands\\
$^3$Institute of Astronomy, University of Cambridge, Madingley
Road, CB3 0HA Cambridge, UK
}

\begin{document}
\bibliographystyle{mn2e.bst}

\date{Accepted 2009 May 28. Received 2009 February 17}

\pagerange{\pageref{firstpage}--\pageref{lastpage}} \pubyear{2009}

\maketitle

\label{firstpage}

\begin{abstract}
Stars on eccentric orbits around a massive black hole (MBH) emit bursts of gravitational waves (GWs) at periapse. Such events may be
directly resolvable in the Galactic centre. However, if the star does
not spiral in, the emitted GWs are not resolvable for extra-galactic
MBHs, but constitute a source of background noise. We estimate the
power spectrum of this extreme mass ratio burst background (EMBB) and
compare it to the anticipated instrumental noise of the {\it Laser
  Interferometer Space Antenna} (LISA). To this end, we model the
regions close to a MBH, accounting for mass-segregation, and for
processes that limit the presence of stars close to the MBH, such as
GW inspiral and hydrodynamical collisions between stars. We find that
the EMBB is dominated by GW bursts from stellar mass black holes, and
the magnitude of the noise spectrum $(f S_{\rm GW})^{1/2}$ is at least
a factor $\sim 10$ smaller than the instrumental noise. As an additional result of our analysis, we show
that LISA is unlikely to detect relativistic bursts in the Galactic
centre.
\end{abstract}

\begin{keywords}
black hole physics  --- stellar dynamics --- gravitational
waves --- Galaxy: centre
\end{keywords}

\section{Introduction}\label{s:intro}

A new opportunity to study stellar processes near massive black holes
(MBHs) arises with the anticipated detection of gravitational waves
(GWs) by the Laser Interferometer Space Antenna (LISA). LISA will be a
space-based detector in orbit around the Sun, consisting of three
satellites five million kilometres apart. It will be sensitive to GWs
in a frequency range $10^{-4}\Hz\lesssim f\lesssim 10^{-2}$ Hz. An important source of GWs for LISA is the inspiral of compact objects
onto MBHs in galactic nuclei (e.g. \citeauthor{Hil95} \citeyear{Hil95}; \citeauthor{Sig97} \citeyear{Sig97};
\citeauthor{Iva02} \citeyear{Iva02}; \citeauthor{Fre03}
\citeyear{Fre03}; \citeauthor{Hop05} \citeyear{Hop05},
\citeyear{Hop06}, \citeyear{Hop06b}; \citeauthor{Ama07}
\citeyear{Ama07}; see \citeauthor{Hop06c}
\citeyear{Hop06c} for a review). These are sources on highly eccentric
orbits with periapses slightly larger than the Schwarzschild radius
$r_{S}=2GM_{\bullet}/c^2$ of the MBH, where $M_{\bullet}$ is the mass
of the MBH. The star dissipates energy due to GW emission, and as a
result spirals in. Such extreme mass ratio inspirals (EMRIs) can be
observed by LISA to cosmological distances if the orbital period of
the star is shorter than $P\lesssim 10^4$ sec (\citeauthor{Fin00}
\citeyear{Fin00}; \citeauthor{Bar04b} \citeyear{Bar04b};
\citeauthor{Gai04} \citeyear{Gai04}; \citeauthor{Gla05}
\citeyear{Gla05}). LISA will detect hundreds to thousands of such
captures over its projected 3-5 yr mission life time \citep{Gai04,
  Gai08}.

For most of the inspiral the emitted GWs are not observable by
LISA. These GWs give rise to a
source of confusion noise, possibly obscuring other types of GW
sources. The shape and overall magnitude of this EMRI background has
been studied by \citet{Bar04b}. They do not study the dynamical
requirements of inspiral, but scale their result with possible EMRI
rates of \citet{Fre01, Fre03}. Therefore noise of stars that do not
eventually spiral in is not included. 

\citet{Rub06} show that stars on long periods of a
few years and nearly radial orbits that carry them near the
Schwarzschild radius of the MBH, emit bursts of GWs that for the Galactic centre
will give a signal to noise ratio larger than 5. Taking into
account processes determining the inner radius of the density profile
as well as mass segregation effects, \citet{Hop07} find that stellar
mass black holes (BHs) have a burst rate of order $1$ yr$^{-1}$, while
the rate is $\lesssim 0.1$ yr$^{-1}$ for main sequence stars (MSs) and
white dwarfs (WDs).

Individual bursts from star-MBH fly-bys may be detected from our own
Galactic centre, but it is unlikely they will be observed from other
galactic nuclei.  However, the accumulation of all bursts of all galaxies
in the universe gives rise to an extreme mass ratio burst
background (EMBB). Since the energy emitted per event
is much higher for inspirals compared to fly-bys, but the event rate
is much lower \citep{Ale03b}, it is not a priori clear which of these
dominates the confuse background. In this paper we study the contribution of these fly-bys to the GW background. 
In \S\ref{s:single} we derive an analytical expression for the event
rate of GW bursts in a galactic nucleus.
This model diverges at several boundaries, and we consider in
\S\ref{sec:processes} what the physical processes are that determine
the range within which our model is valid. Our formalism is similar to
that of the galactic burst rate of \citet{Hop07}. We derive an
expression for the emitted energy spectrum of GW bursts in a single
galactic nucleus in \S\ref{s:em}. A relation for the EMBB from the
accumulation of GW bursts from all redshifts observed by LISA is
derived in \S\ref{s:stoch}. In \S\ref{s:results} we compare the
resulting EMBB to the instrumental noise of LISA \citep{Lar01}, and to other astrophysical backgrounds. We conclude in \S\ref{s:disc} and discuss what our model implies for the possibility of directly detecting relativistic bursts of GWs from our Galactic centre.

\section{The gravitational wave burst rate and spectrum from a single galactic nucleus}\label{s:single}

In this section, we present a simple model for a galactic nucleus and derive the burst rate for that model. We discard parts of phase-space in our model where stars are unlikely to exist due to various processes.

\subsection{The gravitational wave burst rate}

A MBH dominates the potential inside the radius of influence
\begin{eqnarray}\label{e:rh}
r_{\h}={GM_{\bullet}\over \sigma^2},
\end{eqnarray}
where $\sigma$ denotes the stellar velocity dispersion far away from
the MBH. The MBH mass is empirically related to the velocity
dispersion of the bulge by
\begin{eqnarray}
M_{\bullet} = 10^8\ M_{\odot}\ \left( \dfrac{\sigma}{200\
\mathrm{km\ s}^{-1}}\right)^4
\label{eq:msigma}
\end{eqnarray}
(e.g. \citeauthor{Mer01} \citeyear{Mer01}; \citeauthor{Tre02}
\citeyear{Tre02}). Inside the radius of influence, stellar orbits are
assumed to be Keplerian.

Galactic nuclei with $M_{\bullet}\lesssim10^{7}M_{\odot}$ have
relaxation times less than a Hubble time and are dynamically relaxed,
while for very massive MBHs, the distribution will depend on the
initial conditions. However even in this
case the inner region of the cusp may be of power-law form due to the effect caused by a slow growth of a black hole inside a stellar system \citep[see e.g.][]{You80, Qui95}. We will for simplicity assume that our models also
apply for $M_\bullet>10^7M_\odot$. The contribution of such MBHs is
only marginal for the EMBB, and this assumption does not affect our result.

We assume the distribution within $r_h$ to be spherically
symmetric in space and approximately isotropic in velocity space. The
radial density profile is approximated by a power law, $\nu \propto
r^{-\alpha_{\M}}$ (e.g. \citeauthor{Pee72} \citeyear{Pee72}; \citeauthor{Bah76} \citeyear{Bah76}), where the slope $\alpha_{\M}$ depends on the
stellar mass $M_{\star}$; due to mass segregation \citet{Spi87}, larger masses have
larger $\alpha_M$.  The number of stars $n_M(a, J^2)\ d a\ d J^2$ of type $M$ in an
element $(a,\ a+d a), (J^2,\ J^2+d J^2)$ is then given by

\begin{eqnarray}\label{e:naJ2}
n_M(a, J^2)\ d a\ d J^2 = (3-\alpha_{\M})\ {C_{\M}N_{\h}\over r_{\h}}\ \left({a\over
r_{\h}}\right)^{2-\alpha_{\M}}\ {1\over J_{\cc}^2(a)}\ d a\ d J^2,
\end{eqnarray}
where we assume $N_{\h}={\Mbh/\Mo}$. The total number
of stars of type $M$ within $r_{\h}$ is defined as $C_{\M}N_{\h}$. 

We calibrate the parameters $C_{\M}$ using solutions to Fokker-Planck equations applied to the Galactic centre \citep{Ale09}, such that the number of stars within $0.01 \pc$ agrees with those simulations. Since the density distribution of a species $M$ is not a strict power law in the Fokker-Planck models, this implies that our density profile is a good approximation only at small radii. Since the burst rate is dominated at those small radii (see equation [\ref{eq:gamma}]), the small mismatch at larger radii will not strongly affect our conclusions.

A useful parametrisation of an orbit is the frequency
a star would have on a circular orbit with radius of its periapse
$r_{\p}$,
\begin{eqnarray}
\omega_{\p} \equiv \sqrt{\dfrac{GM_{\bullet}}{r_{\p}^3}}.
\label{eq:f}
\end{eqnarray}
Note that for eccentric orbits $\omega_{\p}\gg1/P$ where $P(a)=2\pi
(a^3/GM_{\bullet})^{1/2}$ is the orbital period. The number of stars
$n(a, \omega_{\p})\ d a\ d \omega_{\p}$ in an element $(a,\ a+d a), (\omega_{\p},\ 
\omega_{\p}+d \omega_{\p})$ is then

\begin{eqnarray}
n_M(a,\omega_{\p}) = (3-\alpha_{\M})\ \dfrac{4C_{\M} N_{\h}}{3r_{\h}^2}\ (G\Mbh)^{1/3}\ \left(\dfrac{a}{r_{\h}}\right)^{1-\alpha_{\M}}\ \omega_{\p}^{-5/3},
\label{eq:naf}
\end{eqnarray}
where it was assumed that $e\approx 1$.

A star on a highly eccentric orbit emits almost all of its GW energy
at periapse in a single GW burst. The burst rate per logarithmic unit of the
semi-major axis and frequency at which stars of species $M$ have a
bursting interaction with the MBH is given by

\begin{eqnarray}
\lefteqn{ \dfrac{d ^2\Gamma_{\M}}{d \ln a\ d \ln \omega_{\p}} } \nonumber \\
\lefteqn{ = \frac{1}{P(a)}\ (3-\alpha_{\M})\
\dfrac{4C_{\M} N_{\h}}{3r_{\h}}\ (GM_{\bullet})^{1/3}\
\left(\dfrac{a}{r_{\h}}\right)^{2-\alpha_{\M}}\ \omega_{\p}^{-2/3}}  \\
\lefteqn{= 5.7 \times 10^{-7}\ \mathrm{yr^{-1}}\ (3-\alpha_{\M})\ C_{\M}\ \tilde M\ ^{\alpha_{\M}/2+1/3}\ \tilde a\
^{1/2-\alpha_{\M}}\ \tilde \omega_{\p}\ ^{-2/3},}\nonumber
\label{eq:gamma}
\end{eqnarray}
where we convert to dimensionless units defined as $\tilde{\Mbh} = \Mbh/M_{\circ}$; $\tilde{a}=a/r_{h, \circ}$; $\tilde{\omega_{p}}=\omega_{p}/\omega_{\circ}$, and $M_{\circ}=10^6\Mo$; $r_{h, 0}= GM_{\circ}/\sigma_{\circ}^2\approx1\pc$; $\omega_{\circ}= c^3/(G\M_{\circ})\approx0.2\Hz$.

\subsection{The inner region of the stellar cusp}
\label{sec:processes}
Since we describe the profile of the galactic nucleus by power laws,
integration over orbital space leads to formal divergences, and it is
therefore important to determine the boundaries of validity
of our model. In the next section we consider a number of processes
that can determine the inner edge of the stellar cusp. The burst rate is approximately proportional to $d\Gamma_{\M}/d\ln a \propto
a^{1/2-\alpha_{\M}}$ (see equation [\ref{eq:gamma}]). As typical values for $\alpha$ are in the range
$1.4-2$ \citep[see e.g.][]{Hop06, Fre06, Ale09}, the GW energy spectrum is
dominated by GW bursts from stellar orbits at small radii and
therefore we focus on the stellar dynamics in the vicinity of a MBH.

\subsubsection{ Gravitational wave inspiral}
\label{sec:insp}

As a star loses energy repeatedly with every periapse passage due to
GW emission, it spirals in. Unresolvable EMRIs constitute a background
noise for LISA studied by \citet{Bar04b}. Our study instead focuses
on the subset of extreme mass ratio events that do {\it not} spiral in,
and GW emission from EMRIs must be excluded. We note that the dynamics
of the progenitors of EMRI noise and burst noise is very
different. The part of phase space where stars spiral in is at any
particular time typically not populated, so that, e.g., equation
(\ref{eq:gamma}) does not apply.

The amount of energy $\Delta E_{\GW}$ that is lost in one period is given by \citet{Pet63} as

\begin{eqnarray}
\Delta E_{\GW} & = & \dfrac{8\pi}{5\sqrt2}\ f(e)\ \dfrac{M_{\star}}{M_{\bullet}}\ M_{\star} c^2\ \left (\dfrac{r_{\p}}{r_{\s} } \right )^{-7/2}\nonumber \\
& = &2.8\times10^{49}\ \mathrm{erg}\ \tilde M_{\star}^2\ \tilde M_{\bullet}^{4/3}\ \tilde \omega_{\p}^{7/3},
\label{eq:deltaegw}
\end{eqnarray}
where for highly eccentric orbits $f(e) \approx 0.39$. We define the
characteristic time for inspiral $t_{\ii}$ as the time it takes the
initial orbital energy $\epsilon_0$ to grow formally to infinity,
\begin{eqnarray}
t_{\ii} & = &\int_{\epsilon_0}^{\infty}\dfrac{d \epsilon}{d \epsilon/d t}  \approx \int_{\epsilon_0}^{\infty}\dfrac{d \epsilon}{\Delta
E_{\GW}/P(a)} \nonumber \\ 
&\approx& 2.9\times 10^2\ \mathrm{yr}\ \tilde M_{\star}^{-1}\ \tilde M_{\bullet}^{-5/6}\ \tilde a^{1/2}\
\tilde \omega_{\p} ^{-7/3}.
\label{eq:tinsp}
\end{eqnarray}

While the orbit decays, two body scattering by other stars 
perturbs it, changing the orbital angular momentum by order of itself in a
timescale $t_{\jj}$.  The magnitude and direction of the step in angular
momentum is a random walk process, and typically in a relaxation time
the step is of the size of the circular angular momentum
$J_{\cc}=(GM_{\bullet}a)^{1/2}$, and hence
\begin{eqnarray}
t_{\jj}=\left( \dfrac{J}{J_{\cc}}\right)^2\ t_{\rm relax}.
\label{eq:tj}
\end{eqnarray}
In this expression $t_{\rm relax}$ is the relaxation time,

\begin{eqnarray}
t_{\rm relax}  =  C_{\rm relax}\ \dfrac{\sigma^3}{G^2 \Mo^2 n_{\h}}\ \left({a\over r_{\h}}\right)^{1/2},
\label{eq:trelax}
\end{eqnarray}
where $C_{\rm relax}$ is a numerical constant that also absorbs the
Coulomb logarithm, $\sigma$ is the velocity dispersion far away from
the MBH, and $n_{\h}\propto\Mbh^{-1/2}$ is the density of stars at the
radius of influence.

Due to strong mass-segregation \citep{Ale09, Kes09}, at small radii the relaxation time is
dominated by BH interactions, while at large distances
the MSs dominate the relaxation rate. The true relaxation time is a
combination of the BH and MS relaxation time, and in order to keep the
calculation more tractable, we approximate it by a power law. As the
burst rate is critically sensitive to the inner cut-off of the density
profile, we chose the constant $C_{\rm relax}$ to calibrate the
relaxation time at a small radius of $0.01\pc$ for the model in
\citet{Ale09}. For those values, we find $C_{\rm relax} = 0.02$.

With this scaling the relaxation time and scatter time become

\begin{eqnarray}
&&t_{\rm relax} \!=\! 2.7\ \Gyr\ {C_{\rm relax}\over 0.02}\ \tilde{\Mbh}\ \tilde{a}^{1/2};\nonumber\\
&&t_{\jj}\!=\! 240\ \mathrm{yr}\ {C_{\rm relax}\over 0.02}\ \tilde M_{\bullet}^{4/3}\ \tilde a^{-1/2}\ \tilde \omega_{\p}^{-2/3}.
\label{eq:trelax3}
\end{eqnarray}

If $t_{\jj}>t_{\ii}$, stars spiral in faster than they are
replenished by other stars. In the corresponding region of orbit
space, any bursting star is quickly accreted, so this region is
typically empty (see the appendix for further details). The fact that this part of phase-space is not populated has also consequences for the possibility of detecting relativistic bursts in the Galactic centre; see \S\ref{s:disc} for further details. Solving $t_{\jj}=t_{\ii}$ for $\omega_{\p}$ gives an inner cut-off of

\begin{eqnarray}
\tilde \omega_{\p} <\tilde \omega_{\ii} \equiv 1.1\ (\dfrac{C_{relax}}{0.02})^{-3/5}\ \tilde{\Ms}^{-3/5}\ \tilde{\Mbh}^{-13/10}\ \tilde a^{3/5}. 
\label{eq:critinsp}
\end{eqnarray}

We assume that stars with $\tilde \omega_{\p}> \tilde \omega_{\ii}$ always spiral in
and we neglect these stars (noise for these stars was accounted for by
\citeauthor{Bar04b} \citeyear{Bar04b}), while stars with $\tilde \omega_{\p} < \tilde \omega_{\ii}$ never spiral in, and we take these stars into
account as bursting sources.

\subsubsection{The direct capture loss-cone}
All stars with an angular momentum less than the last stable orbit
$J_{\rm LSO}=4GM_{\bullet}/c$ (the ``loss-cone'') are pulled into the
MBH. Since a star in this region of orbit space is removed in a
dynamical time, we do not
consider stars in this region. Therefore stars only contribute GWs to
the EMBB if

\begin{eqnarray}
\tilde \omega_{\p} < \tilde \omega_{lc} \equiv 4.41 \times 10^{-2}\ \tilde M_{\bullet}^{-1}.
\label{eq:flso}
\end{eqnarray}

In reality the empty region inside the loss-cone will also affect the
regions nearby in angular momentum space \citep{Lig77}, although this
is partially erased by resonant relaxation (\citeauthor{Rau96}
\citeyear{Rau96}; \citeauthor{Rau98} \citeyear{Rau98}). Here we assume
that orbits outside the loss-cone are fully populated.

\subsubsection{The tidal disruption loss-cone}
Tidal effects can play a role when the star is not compact. The tidal
forces from the MBH on the MS disrupt those MSs whose orbits carry
them within the tidal radius $r_{\ttt}\approx
(2M_{\bullet}/M_{\star})^{1/3}R_{\star}$ of the MBH, where $R_{\star}$ denotes the radius of the MS. This
effect leads to a depletion of MSs when $r<r_{\ttt}$, called the
loss-cone for tidal disruptions. We only consider MSs if

\begin{eqnarray}
\tilde \omega_{\p}< \tilde \omega_{\ttt} \equiv 2.18 \times 10^{-3}\  \left({M_{\star}\over \Mo}\right)^{1/2}\ \left({ R_{\star}\over \Ro}\right)^{-3/2}.
\label{eq:ftidal}
\end{eqnarray}
Note that the break up frequency $\tilde
\omega_{\ttt}$ depends solely on the characteristics of the MS.

\subsubsection{Hydrodynamical collisions}
\label{sec:coll}
Close to a MBH, MSs are likely to undergo multiple collisions in their
lifetime \citep[see e.g.][]{Spi66}. The rate
$\Gamma_{\rm coll}$ at which a MS with radius $R_{\star}$ has grazing
collisions can be estimated by
\begin{eqnarray}
\Gamma_{\rm coll}=n\ \sigma_{\mathrm{cs}}\ v= \frac{3-\alpha_{\MS}}{4\pi}\ \frac{N_{\h}}{r_{\h}^3}\ \left(\frac{a}{r_{\h}}\right)^{-\alpha_{\MS}}\ \pi R_{\star}^2\ \sqrt{\frac{GM_{\bullet}}{a}}.
\label{eq:collrate}
\end{eqnarray}

Multiple collisions can lead to the disruption of the
MS. \citet{Fre05} showed that on average a MS is disrupted after
$N_{\rm coll} \approx 30$ collisions. This implies that MSs are saved
from disruption by collisions within a Hubble time, if their distance
to the MBH is larger than
\begin{eqnarray}
\tilde a > \tilde a_{\rm coll} \equiv 1.6\times10^{-3}\ \left( {30\over N_{\rm coll}}\right)^{10/19} \left({R_{\star}\over \Ro}\right) ^{20/19}\ \tilde M_{\bullet}^{7/19},
\label{eq:acoll}
\end{eqnarray}
where it was assumed that $\alpha_{\MS} = 1.4$ as in our main model. 
MSs are contributing GWs to the EMBB if their orbits obey $\tilde a >
\tilde a_{\rm coll}$.

\subsection{Model of a galactic nucleus}
\label{sec:mainmodel}

To summarise, the method to compute the GW burst rate of a single
galactic nucleus is as follows. We consider four distinct species of
stars, that is $10 M_{\odot}$ BHs, $0.6 M_{\odot}$ WDs, $1.4
M_{\odot}$ neutron stars (NSs), and $1 M_{\odot}$ MSs. Their density
profile is given by a slope of $\alpha_{\mathrm{BH}}=2$,
$\alpha_{\mathrm{WD}}=1.4$, $\alpha_{\mathrm{NS}}=1.5$,
$\alpha_{\MS}=1.4$ \citep{Ale09}. We assume
that the enclosed number of MS stars within $r_h$ is
$N_{\h}=M_{\bullet}/M_{\odot}$. We calibrate our model such that for the Galactic centre,  the number of stars within $0.01 \pc$ is equal to that found in \citet{Ale09}, to find $C_{\mathrm{BH}}=8\times 10^{-3}$,
$C_{\mathrm{WD}}=0.09$, $C_{\mathrm{NS}}=0.01$, $C_{\MS}=1$. From
their models, we find that the relaxation time at $0.01\pc$ in the
Galactic centre is $0.78\Gyr$, leading to a calibration factor of
$C_{\rm relax}=0.02$.

We also consider a model suggested\footnote{\citet{OLe08} consider a wide range of models,
  including models with steeper mass functions. We will refer to the
  model considered here as the ``\citet{OLe08} model'' for brevity. \citet{OLe08} assume a range of BH masses, which we collect here into a single averaged BH mass.}
by \citet{OLe08}. In this model, the density of the BHs at the radius of influence is higher by a
factor 10 at $r_h$ compared to the main model, and the masses of the
BHs are higher. Such models may be relevant for Galactic nuclei
because of empirical hints that the initial mass function is much
flatter than usual the Galactic centre (see e.g. \citeauthor{Nay05}
\citeyear{Nay05}; \citeauthor{Man07} \citeyear{Man07}). We note that
there are no dynamical constraints on the amount or mass of BHs in the
Galactic centre \citep[e.g.][]{Mou05, Ghe08}.
Since in systems with many, more massive BHs mass segregation is less
pronounced, the scaling for the density profile is not
straightforward; for example, the
number of BHs enclosed within $0.01\pc$ is only a factor $\sim2$
higher than in our main model. Using the relaxation times given in
\citet{OLe08}, we find that for this model $M_{\rm BH}=18\Mo$; $C_M=1.6\times10^{-2}$; $\alpha_{\rm BH}=2$; and $C_{\rm
  relax}=3\times10^{-3}$.

\begin{figure*}
\begin{center}
\begin{tabular}{c c}
\scalebox{0.35}{\includegraphics{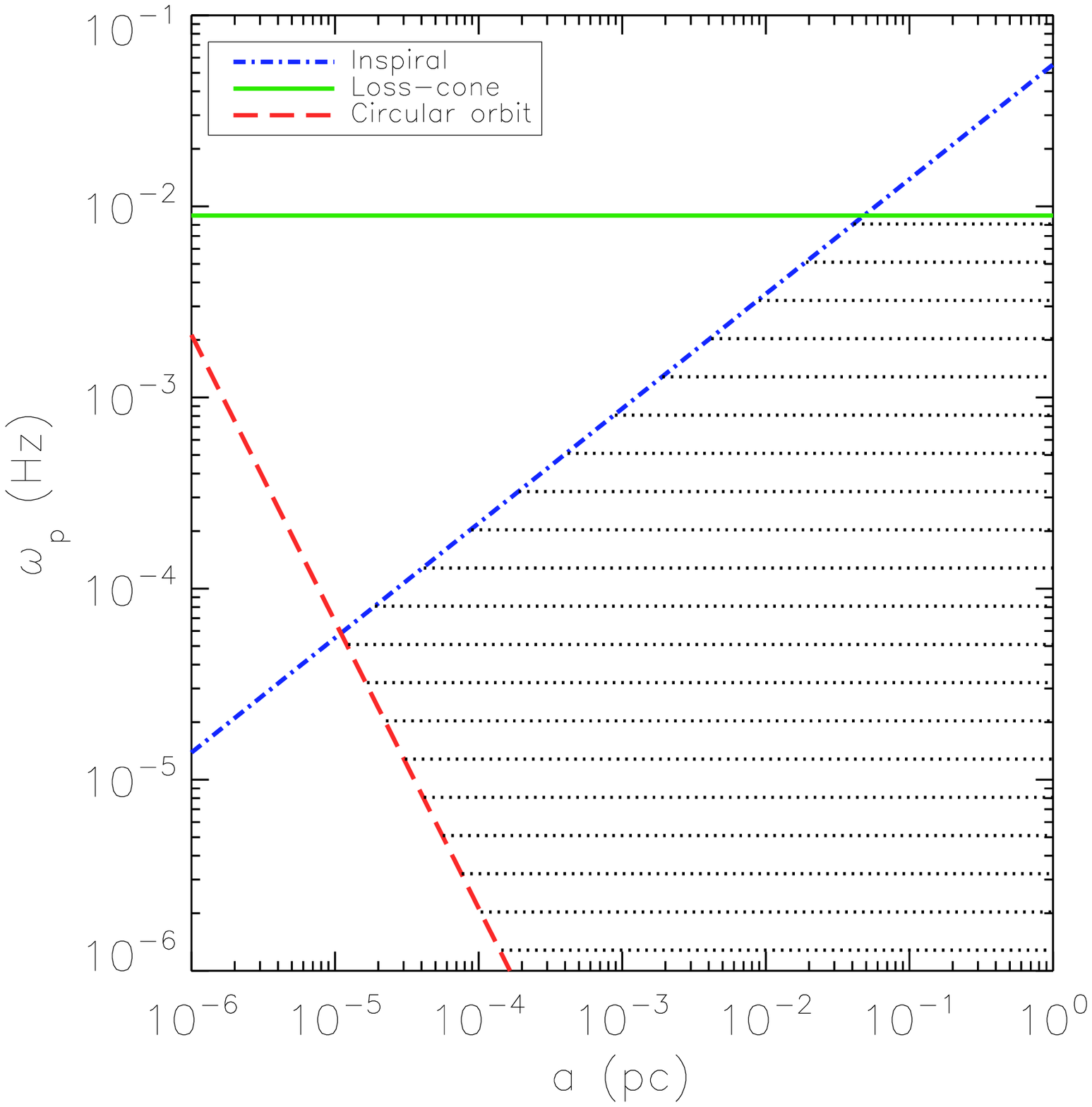}}
&
\scalebox{0.35}{\includegraphics{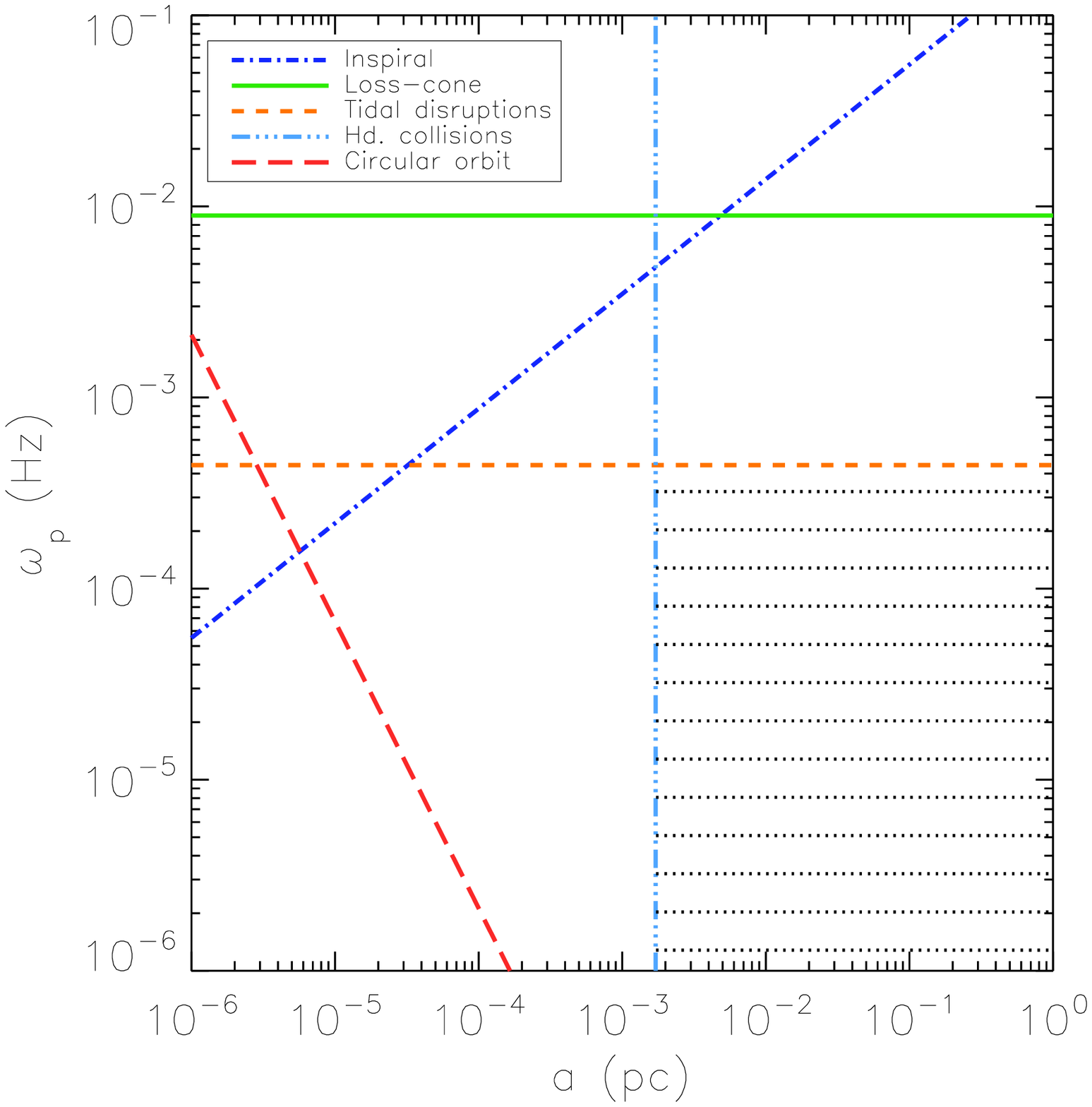}} \\
\small \emph{(a)} & \small \emph{(b)}\\
\end{tabular}
\caption{ Boundaries in orbit space around a MBH of $M_{\bullet}=10^6
  M_{\odot}$ for a BH of $M_{\star}=10 M_{\odot}$ in
  Fig. \ref{fig:plotcut}a and of a MS of $M_{\star}=1 M_{\odot}$ in
  Fig. \ref{fig:plotcut}b. Stellar orbits are populated in the shaded
  area only. The figures are representative of all relevant stars and
  MBHs where the cuts scale according to Eq. (\ref{eq:critinsp}),
  (\ref{eq:flso}), (\ref{eq:ftidal}) and (\ref{eq:acoll}). The red
  line indicates the boundary in orbit space resulting from the
  condition $r_{\p}\leqslant a$. The result is not very sensitive
  to the inner cutoff; imposing a hard cutoff at $10^{-4}$ pc in
  figure (a) decreases $(f S_{\GW})^{1/2}$ by $\sim 50$ per cent. }
\label{fig:plotcut}
\end{center}
\end{figure*}

When using these models, we exclude certain orbits due to the four processes described in the
previous section. Stars are only contributing GWs if the inequalities
(\ref{eq:critinsp}), (\ref{eq:flso}), (\ref{eq:ftidal}) and
(\ref{eq:acoll}) are satisfied.  As an illustration we show the orbit
space around a MBH of $M_{\bullet}=10^6\ M_{\odot}$ for a BH of
$M_{\star}=10\ M_{\odot}$ in figure (\ref{fig:plotcut}a) and for a MS of
$M_{\star}=1\ M_{\odot}$ in figure (\ref{fig:plotcut}b). The boundaries
of the populated parts of the orbit space of compact objects are
caused by two processes; the inspiral process and the loss-cone. The
cut-offs through the orbit space of MSs are caused by all four
processes of \S\ref{sec:processes}. We note that the processes that
limit the GW burst rate depend on the MBH mass.

\section{Gravitational wave background}\label{s:em}
\subsection{Gravitational wave energy in a single burst}
\label{sec:singleenergy}
For an eccentric Newtonian orbit, the amount of energy emitted in GWs
in one period $\Delta E_{\GW}$ [see Eq. (\ref{eq:deltaegw})] and the
distribution of this energy over frequencies $dQ/df$ is studied by
\citet{Pet63}. The emitted power peaks at a frequency $f$ that is
twice the circular frequency at periapse $\omega_{\p}$. To simplify
our analysis we assume that all energy $\Delta E_{\GW}$ is emitted at
$f=2\omega_{\p}$, therefore
\begin{eqnarray}
\dfrac{d Q}{d f} = \Delta E_{\GW}\ \delta(f-2\omega_{\p}).
\label{eq:dQdw}
\end{eqnarray}
In reality, the spectrum is quite broad, but the exact shape of the spectrum is not important, since we sum over
a very large number of bursts.

\subsection{Gravitational wave energy density}
\label{sec:numberdensityMBH}
We next integrate this spectrum over $M_{\bullet}$ with
the space number density of MBHs of \citet{All02}, which we assume to
be constant throughout the history of the Universe,
\begin{eqnarray}
\dfrac{d n_{\bullet}}{d M_{\bullet}}= \dfrac{10^7}{M_{\bullet}}\ \mathrm{Gpc}^{-3}\quad\quad \left(10^5\ M_{\odot}<M_{\bullet}<10^8\ M_{\odot}\right).
\label{eq:dndm}
\end{eqnarray}
The upper boundary
comes from a sharp drop in the observed MBH number density around
$M_{\bullet}\approx 10^8\ M_{\odot}$. Due to a lack of information on
MBHs of $M_{\bullet}<10^6\ M_{\odot}$, the lower boundary is
arbitrarily chosen to be $10^5\ M_{\odot}$. Although at frequencies between $10^{-3}$ and $10^{-2}$ Hz, MBHs of the lowest masses contribute most to the EMBB, decreasing the lower bound only mildly affects the EMBB. 
The energy density of GW
bursts emitted by stars of species $M$ is
\begin{eqnarray}
\dot{\mathcal{E}}(f)= \int\limits_{\mathcal{V}} \dfrac{d ^2\Gamma_{\M}}{d a\ d \omega_{\p}}\ \Delta E_{\GW}\ \delta(f-2\omega_{\p})\ \dfrac{d n_{\bullet}}{d M_{\bullet}}\ d M_{\bullet}\ d \omega_{\p}\  d a.
\label{eq:espec}
\end{eqnarray}

\subsection{Observed gravitational wave energy density}\label{s:stoch}
The GW spectrum observed by LISA is the accumulation of the GW
radiation of all galaxies in the universe. The emitted spectrum $\dot{\mathcal{E}}(f_{\rm em})df_{\rm em}$ is the
rate per unit proper time and per unit co-moving volume at which GW
energy in the frequency range $(f_{\rm em}, f_{\rm em} + d f_{\rm em})$ are
emitted. The total GW energy density $(d\rho/df) df$ today with frequencies $f=f_{\rm em}/(1+z)$ in the range $(f, f + d f)$ is then

\begin{eqnarray}
{d \rho\over d f} df  &=& d f_{\rm em}\int dz \dfrac{d t}{d z} \frac{\dot{\mathcal{E}}(f_{\rm em})}{1+z}  \nonumber \\
& = &
d f\int dz \dfrac{d t}{d z}\dot{\mathcal{E}}[f(1+z)].
\label{eq:z}
\end{eqnarray}
For $0<z<2$, we follow
\citet{Bar04b} in approximating $t\approx t_0\ (1+z)^{-1.18}$, where
a spatially flat Friedmann-Lema\^itre-Robertson-Walker Universe is
assumed with $\Omega_{\Lambda}=0.70$ and $\Omega_{\M}=0.30$ and the
Universe's current age is $t_0 = 0.964H_0^{-1} = 1.39 \times 10^{10}
h_{70}^{-1}$ yr. For sources in the range $0<z<2$, this is accurate to
within $\sim$3 per cent. The approximation is justified since the EMBB
is dominated by the nearby universe and contributions from $1<z<2$
accounts for just 5-8 per cent of that of $0<z<2$.

\subsection{Gravitational wave background for LISA}
The energy density of an isotropic background of individually
unresolvable GW sources is related to the spectral density in the LISA
detector by $S_{\GW}(f) = 4G/(\pi c^2 f^{2})\ d \rho_{\M}/d f$ \citep{Bar04b}, or

\begin{eqnarray}
S_{\GW}(f) =  \dfrac{4G}{\pi c^2 f^{2}}\  \int \limits_{\mathcal{V}}
 & d z\  d M_{\bullet}\ d\omega_{\p}\ d a\ \dfrac{d t}{d z} \dfrac{d n_{\bullet}}{d M_{\bullet}}  \dfrac{d^2\Gamma_{\M}}{d a\ d \omega_{\p}}\  \nonumber \\ 
& \times\ \Delta
E_{\GW}\ \delta[f(1+z)-2\omega_{\p}] 
\end{eqnarray}
where  equations (\ref{eq:espec}) and (\ref{eq:z}) were used. The integral is taken over $\mathcal{V}$ representing that part of
orbit space that is populated according to \S\ref{sec:processes}. 

\section{Results}
\label{s:results}

\begin{figure}
\centering
\includegraphics[angle=0,scale=.4]{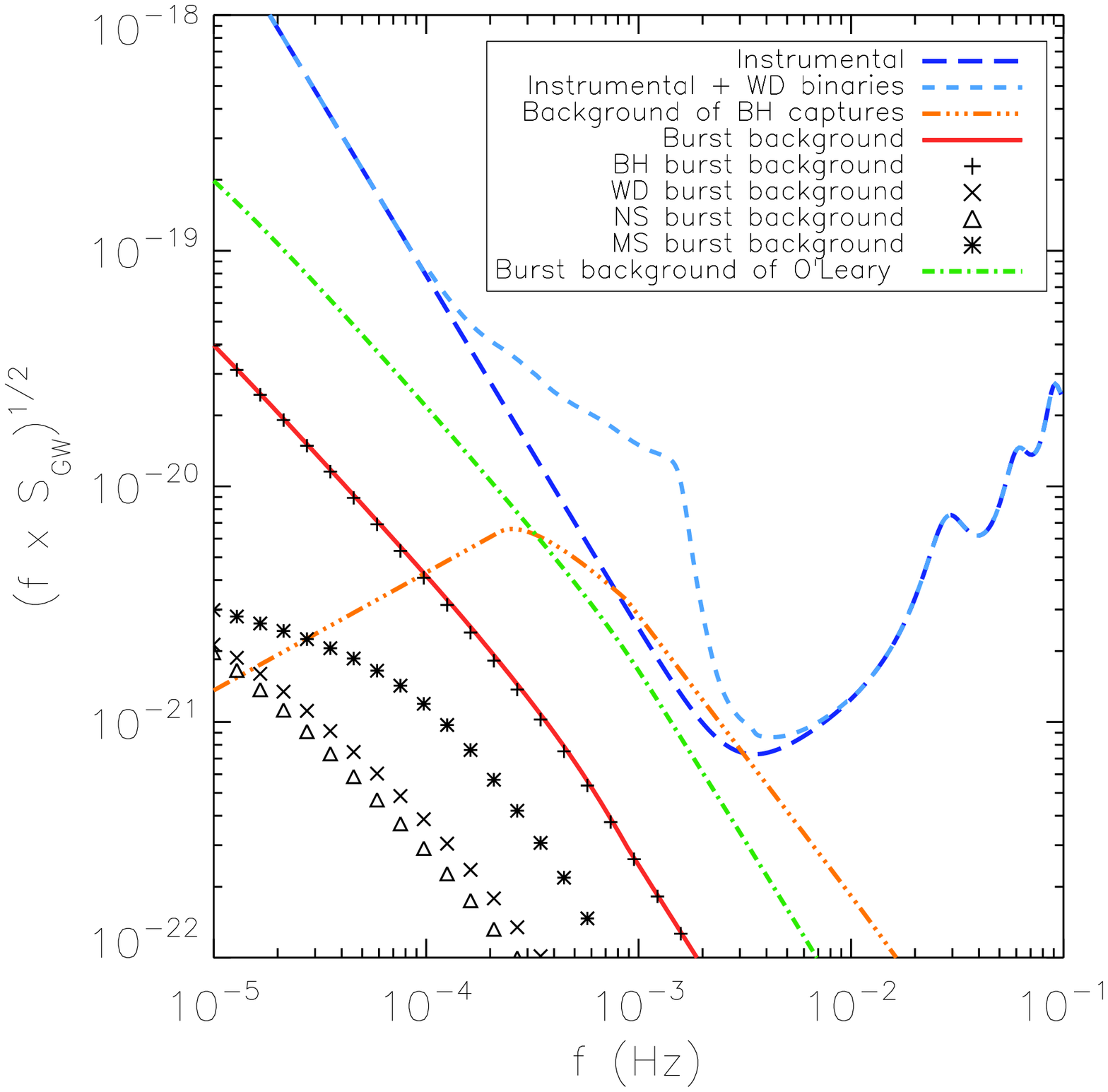}
\caption{Comparison between the EMBB, instrumental noise of LISA (\citeauthor{Lar01} \citeyear{Lar01}), the galactic WD binary background (\citeauthor{Lar01} \citeyear{Lar01}) and the BH EMRI (inspiral) background (\citeauthor{Bar04b} \citeyear{Bar04b}) with an inspiral rate of $10^{-7}{\rm yr^{-1}}$. The EMBB is smaller than the other types of GW noise. The EMBB due to only BHs, WDs, NSs or MSs is shown in symbols and contributions from BHs dominate the spectrum. The EMBB of the \citet{OLe08} model is also shown. Even though it is significantly higher than the EMBB in our preferred model, it is still lower than the instrumental noise of LISA.}
\label{fig:zplota}
\end{figure}
For our main model, the resulting EMBB $(f S_{\GW})^{1/2}$ of the four
species of stars is shown in figure (\ref{fig:zplota}). The EMBB is
dominated by contributions from BHs. This is mainly due to two
reasons. First, a single burst emitted by a BH is more energetic than
a burst emitted by the other species of stars in a similar
orbit. Second, the BH distribution is steeper due to mass-segregation
which leads to a higher burst rate, see Eq. (\ref{eq:gamma}). Note
that the EMBB due to MSs is cut off at $\sim 10^{-3}$ Hz, because
tidal effects prohibit the existence of non-compact objects at small
distances from the MBH. 

The EMRI background is caused by BHs
that spiral in to the MBH. These sources are excluded in our
calculations by eliminating inspiral orbits, see \S\ref{sec:insp}.  The expression
for the EMRI background was taken from \citet{Bar04b} and scales with
the inspiral rate. From \citet{Hop06b} we take an inspiral rate of $1
\times 10^{-7}$ yr$^{-1}$ leading to a EMRI background at a level
comparable to the instrumental noise of LISA. 

In our favoured model the EMBB is well below the instrumental noise of
LISA. At closest approach the EMBB $(f S_{\GW})^{1/2}$ is a factor
$\sim$ 10 below the instrumental noise of LISA, and a factor $\sim$ 19
lower than the confusion noise of Galactic WD binaries. The EMBB is
also lower than the EMRI background in the interesting frequency range
where the noise from EMRIs is stronger than that of the instrument
itself.

The \citet{OLe08} model is also displayed in figure (\ref{fig:zplota}). In this model there are more, and more
massive BHs, which raises the EMBB due to an increase in the number of
BH bursts and an increase in the relaxation rate protecting stars from
inspiral. Figure (\ref{fig:zplota}) shows the EMBB is raised by a factor
$\sim 6$ on average with respect to our main model. However, even with
this stellar distribution the EMBB is smaller than the instrumental
noise of LISA by a factor $\sim 1.5$, and a factor $\sim 3$ including
the noise of galactic WD binaries. Overall the sensitivity of the EMBB on the individual model parameters is not strong enough to raise the EMBB above the instrumental noise of LISA, without increasing $M_{\mathrm{BH}}$ by a factor 10 and other parameters by even more. 

\section{Summary and discussion}\label{s:disc}
Stars whose orbits carry them near the Schwarzschild radius of a MBH
emit gravitational waves in the LISA frequency band. If the orbit of
the star is very eccentric, the star emits bursts of GWs each
peri-centre passage without necessarily spiraling in. Most extra-galactic bursts
are not individually resolvable and hence will constitute a
gravitational wave background.

For our favoured model the EMBB is dominated by GW bursts from BHs,
and is a factor $\sim10$ lower than the instrumental
noise of LISA. Including the WD Galactic background increases the
insignificance of the EMBB to a factor $\sim19$. Even for the
\citet{OLe08} model, which has a much flatter mass function, the
EMBB is smaller than the instrumental noise of LISA by a factor $\sim
1.5$ (or a factor $\sim3$ including the WD Galactic background). The
EMBB is also well below the EMRI background (\citeauthor{Bar04b}
\citeyear{Bar04b}). We conclude that the detection of GWs by LISA will
not be hindered by a background of bursting sources.

As an additional application of our model, we revisit the possibility of direct detection of GW bursts in the Galactic centre, due to stars on relativistic orbits.
\citet{Yun08} have investigated the impact of including relativistic
corrections to the description of the star's trajectory. The degree to
which the relativistic corrections are important depends on the star's
orbit and is largest for small peri-centre distances. \citet{Yun08}
find that orbits with peri-centre velocities $|v_{\p}|>0.25c$ account
for approximately half of the events within the orbit space considered
by \citet{Rub06} around a MBH of $3.7\times10^6\ M_{\odot}$. However, \citet{Rub06}
incorporate tight, possibly relativistic orbits that are unpopulated
according to our assumptions of the boundaries in orbitspace of
\S\ref{sec:processes} (see also the appendix).  Assuming our main model of a galactic nucleus,
a similar MBH mass as \citet{Rub06} and integrating over all frequencies that
can be detected by LISA (See Eq. (4) in \citet{Hop07} with a signal to
noise of 5), relativistic orbits account for $\sim 1$ per cent of all
events within the orbit space considered by our main model. We
conclude that LISA is unlikely to detect relativistic bursts from the
Galactic centre.

\section*{Acknowledgements}
CH is supported by a Veni scholarship from the
 Netherlands Organisation for Scientific Research (NWO).

\appendix
\section{Depletion of phase space due to gravitational wave inspiral}\label{s:appA}

Throughout this paper, we have not considered the contribution of stars in the region where $t_J>t_i$ (see \S\ref{sec:insp}). Stars in this region {\it do} contribute to the EMBB: these stars spiral in and become eventually directly resolvable EMRIs. Before they can be resolved, however, they contribute to a stochastic background. This contribution to the EMBB was studied by \citet{Bar04b}, and we find that it is dominant over the contribution of fly-bys (see figure \ref{fig:plotcut}).

We stress that our analysis as presented in this paper does not apply to the region $t_J>t_i$, because orbits in this region are less likely to be populated. In this appendix we analyse a toy model of Monte Carlo (MC) simulations based on the method by \citet{Sha78} to show this. For details on the method, we refer to \citet{Sha78} and \citet{Hop09}. The model does not include mass-segregation, and is not intended to faithfully represent a galactic nucleus, but is instead used to highlight the difference in the dynamics in the two regimes delineated by $t_j=t_i$.

Consider the following simple dynamical model:

{\it Energy diffusion ---} Let the (scaled) energy $x$ perform a random walk such that the diffusion coefficient is

\begin{equation}
D_{xx} = 85x^{9/4}.
\end{equation}
The form of this coefficient is such that it reproduces the \citet{Bah76} solution with distribution function $f(x)\propto x^{1/4}$; the prefactor is for consistency with \citet{Hop09}. (Here positive energy implies that the star is bound to the MBH).

{\it Angular momentum diffusion ---} Let the angular momentum $j$ be scaled by the circular angular momentum, such that by definition $0<j<1$. Let $j$ perform a random walk such that the diffusion coefficient is (see \citet{Hop09} for the prefactor)

\begin{equation}
D_{jj} = 6.6x^{1/4}.
\end{equation}

{\it Boundary conditions ---} Let the actual angular momentum of the loss-cone be independent of energy, such that if it is scaled by the circular angular momentum, it becomes

\begin{equation}
j_{lc}(x) = \left({x\over x_{\rm max}}\right)^{1/2},
\end{equation}
with $x_{\rm max}$ some maximal energy.

{\it Energy loss by GWs ---} Let the loss of energy per unit time be

\begin{equation}
\dot{x}_{\rm GW} = x^{3/2}\left({j\over j_{lc}(x)}\right)^{-7}\tau^{-1},
\end{equation}
which has a scaling such as that in GW emission, and the constant $\tau$ is chosen here to be $\tau=10^{-6}$.

This leads to the following dynamical equations

\begin{equation}
j_{t+\delta t} = j_{t} + r \sqrt{D_{jj} \delta t};
\end{equation}

\begin{equation}
x_{t+\delta t} = x_{t} + r \sqrt{D_{xx} \delta t} + \dot{x}_{\rm GW}\delta t,
\end{equation}
where in these equations $r$ is a normally distributed random variable with mean zero and unit variance. The three terms in the dynamical equations give rise to three time-scales: the time scale for changes in $j$ of order of itself due to scattering

\begin{equation}
t_j = D_{jj}^{-1}j^2 = (1/6.6)x^{-1/4}j^2;
\end{equation}
the time scale for changes in $x$ of order of itself due to scattering
\begin{equation}
t_x = D_{xx}^{-1}x^2 = (1/85)x^{-1/4};
\end{equation}
and the time for changes in $x$ of order of itself due to GW emission
\begin{equation}
t_{i} = {x\over \dot{x}_{\rm GW}} = x^{-4}j^{7}x_{\rm max}^{7/2}\tau.
\end{equation}

\begin{figure}
	\begin{center}
			\includegraphics[height=80 mm,angle=270 ]{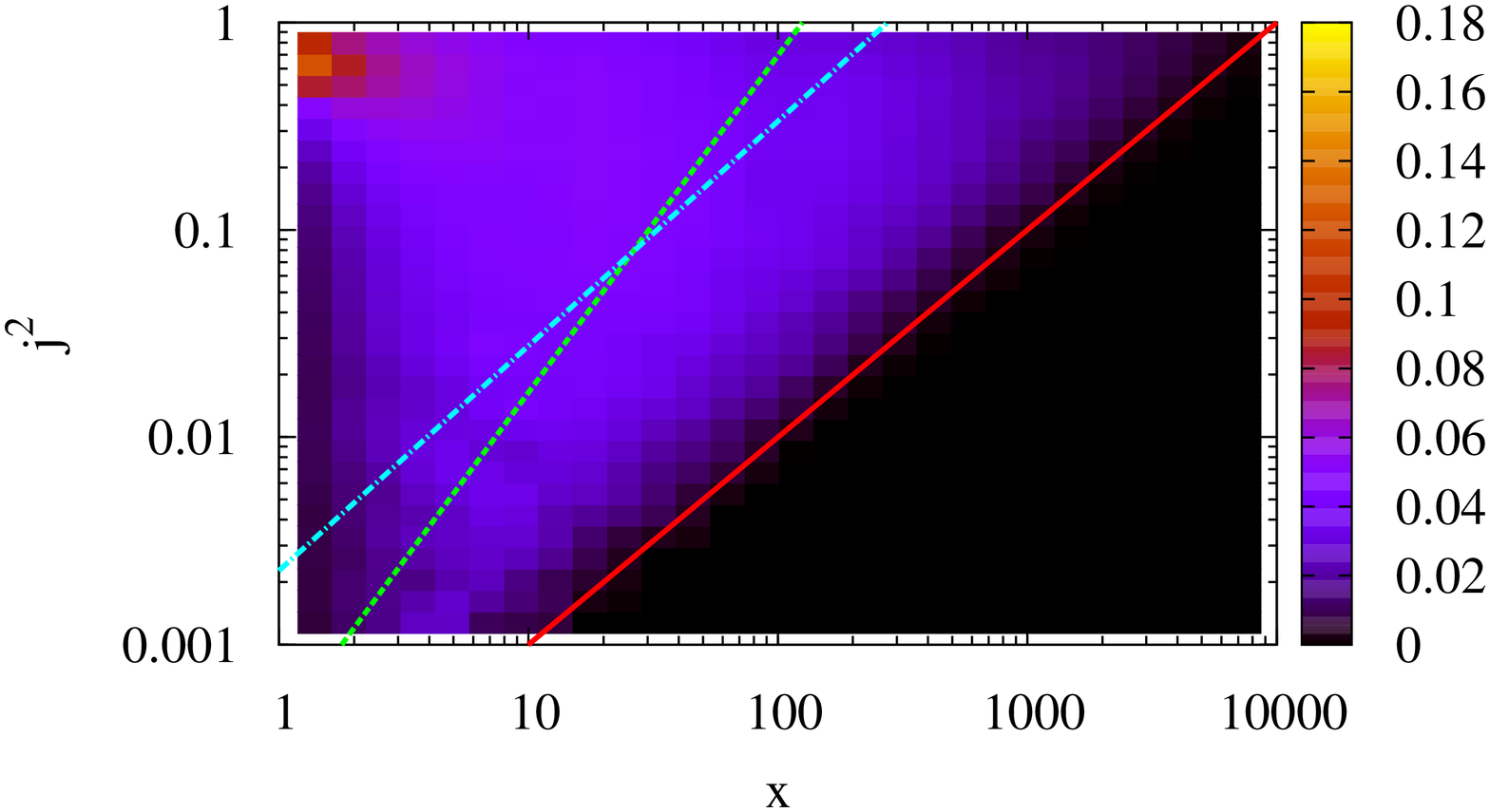}
		\caption{Distribution $f(x, j^2)$ of stars without GW emission (arbitrary scale). The red solid line denotes the loss-cone, the blue dotted line delineates $t_j=t_{i}$, and the green dashed line delineates $t_x=t_{i}$. \label{f:f1}}
	\end{center}
\end{figure}

\begin{figure}
	\begin{center}
			\includegraphics[height=80 mm,angle=270 ]{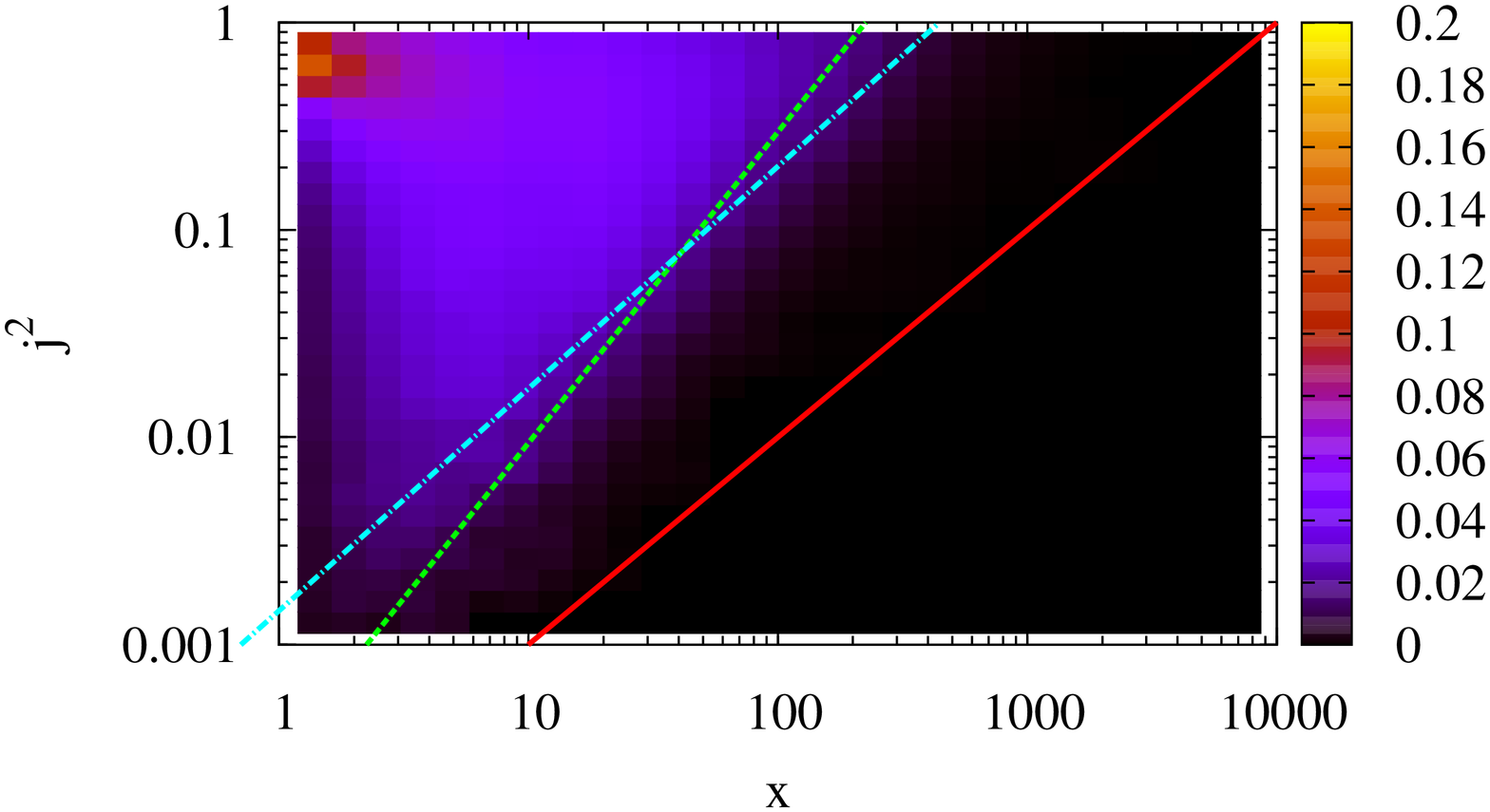}
		\caption{Same as previous figure, but now with energy losses to GWs. Comparison of the two figures shows that the distribution is depleted in presence of GW emission, approximately in the region where $t_{i}<\min(t_j, t_{x})\sim t_j$. Note that some stars do remain within the $t_{i}<\min(t_j, t_{x})$ region; these are inspiraling stars, and their contribution to the EMBB was accounted for by \citet{Bar04b}.  \label{f:f2}}
	\end{center}
\end{figure}

We simulate the dynamical process described above with and without GW emission by means of a MC simulation with about $10^6$ particles. We follow the \citet{Sha78} cloning scheme in order to resolve the distribution function over several orders of magnitude. We define the  distribution $f(x, j^2)$ as

\begin{equation}
f(x, j^2) \equiv {x^{5/4}\over j^2}{d^2 N(x, j^2)\over d\ln x d\ln j^2},
\end{equation}
where $N$ is the normalised number of particles and the prefactor $x^{5/4}/j^2$ was added in order to divide out the dependence on energy and angular momentum if the distribution is isotropic and there are no GWs. We show the resulting normalised steady state distribution in figure (\ref{f:f1}) for the case without GW emission and figure (\ref{f:f2}) for the case with GW emission. The figures clearly shows that the region where $t_i<t_j$ is depleted if there are energy losses to GWs.

\bibliographystyle{mn2e}

\begin{thebibliography}{}

\bibitem[\protect\citeauthoryear{{Alexander} \& {Hopman}}{{Alexander} \&
 {Hopman}}{2003}]{Ale03b}
{Alexander} T.,  {Hopman} C.,  2003, \apjl, 590, L29

\bibitem[\protect\citeauthoryear{{Alexander} \& {Hopman}}{{Alexander} \&
  {Hopman}}{2009}]{Ale09}
{Alexander} T.,  {Hopman} C.,  2009, ArXiv e-prints, 697, 1861

\bibitem[\protect\citeauthoryear{{Allen} \& {Richstone}}{{Allen} \& {Richstone}}{2002}]{All02}
{Allen} M.~C.,  {Richstone} D.,  2002, \aj, 124, 3035

\bibitem[\protect\citeauthoryear{{Amaro-Seoane}, {Gair}, {Freitag}, {Miller},
 {Mandel}, {Cutler} \& {Babak}}{{Amaro-Seoane} et~al.}{2007}]{Ama07}
{Amaro-Seoane}, P., {Gair}, J.~R., {Freitag}, M., {Miller}, M.~C., {Mandel}, I., {Cutler}, C.~J., {Babak}, S., 2007, Clas. Quantum Gravity,
 24, 113

\bibitem[\protect\citeauthoryear{{Bahcall} \& {Wolf}}{{Bahcall} \& {Wolf}}{1976}]{Bah76} {Bahcall} J.~N.,  {Wolf} R.~A.,  1976, \apj, 209, 214

\bibitem[\protect\citeauthoryear{{Bahcall} \& {Wolf}}{{Bahcall} \&
 {Wolf}}{1977}]{Bah77}
{Bahcall} J.~N.,  {Wolf} R.~A.,  1977, \apj, 216, 883

\bibitem[\protect\citeauthoryear{{Barack} \& {Cutler}}{{Barack} \&
 {Cutler}}{2004}]{Bar04b}
{Barack} L.,  {Cutler} C.,  2004, \prd, 70, 122002

{Cohn} H.,  {Kulsrud} R.~M.,  1978, \apj, 226, 1087
\bibitem[\protect\citeauthoryear{{Finn} \& {Thorne}}{{Finn} \&
 {Thorne}}{2000}]{Fin00}
{Finn} L.~S.,  {Thorne} K.~S.,  2000, \prd, 62, 124021

{Frank} J.,  {Rees} M.~J.,  1976, \mnras, 176, 633

\bibitem[\protect\citeauthoryear{{Freitag}}{{Freitag}}{2001}]{Fre01}
{Freitag} M.,  2001, Clas. Quantum Gravity, 18, 4033
\bibitem[\protect\citeauthoryear{{Freitag}}{{Freitag}}{2003}]{Fre03}
{Freitag} M.,  2003, \apjl, 583, L21
\bibitem[\protect\citeauthoryear{{Freitag} \& {Benz}}{{Freitag} \&
 {Benz}}{2005}]{Fre05}
{Freitag} M.,  {Benz} W.,  2005, \mnras, pp 245
\bibitem[\protect\citeauthoryear{{Freitag}, {Amaro-Seoane} \&
 {Kalogera}}{{Freitag} et~al.}{2006}]{Fre06}
{Freitag} M.,  {Amaro-Seoane} P.,    {Kalogera} V.,  2006, \apj, 649, 91

\bibitem[\protect\citeauthoryear{{Gair}, {Barack}, {Creighton}, {Cutler},
 {Larson}, {Phinney} \& {Vallisneri}}{{Gair} et~al.}{2004}]{Gai04}
{Gair} J.~R.,  {Barack} L.,  {Creighton} T.,  {Cutler} C.,  {Larson} S.~L.,
 {Phinney} E.~S.,    {Vallisneri} M.,  2004, Clas. Quantum Gravity,
 21, 1595

\bibitem[\protect\citeauthoryear{{Gair}}{{Gair}}{2008}]{Gai08}
{Gair} J.~R.,  2008, ArXiv e-prints (arXiv: 0811.0188) 


\bibitem[\protect\citeauthoryear{{Ghez}, {Salim}, {Weinberg}, {Lu}, {Do},
  {Dunn}, {Matthews}, {Morris}, {Yelda} \& {Becklin}}{{Ghez}
  et~al.}{2008}]{Ghe08}{Ghez} A.~M., et~al., 2008, IAU symposium, 248, 52

\bibitem[\protect\citeauthoryear{{Glampedakis}}{{Glampedakis}}{2005}]{Gla05}
{Glampedakis} K.,  2005, Clas. Quantum Gravity, 22, 605


\bibitem[\protect\citeauthoryear{{Hils} \& {Bender}}{{Hils} \&
 {Bender}}{1995}]{Hil95}
{Hils} D.,  {Bender} P.~L.,  1995, \apjl, 445, L7

\bibitem[\protect\citeauthoryear{{Hopman} \& {Alexander}}{{Hopman} \&
 {Alexander}}{2005}]{Hop05}
{Hopman} C.,  {Alexander} T.,  2005, \apj, 629, 362
\bibitem[\protect\citeauthoryear{{Hopman} \& {Alexander}}{{Hopman} \&
 {Alexander}}{2006a}]{Hop06}
{Hopman} C.,  {Alexander} T.,  2006a, \apj, 645, 1152
\bibitem[\protect\citeauthoryear{{Hopman} \& {Alexander}}{{Hopman} \&
 {Alexander}}{2006b}]{Hop06b}
{Hopman} C.,  {Alexander} T.,  2006b, \apjl, 645, L133
\bibitem[\protect\citeauthoryear{{Hopman}}{{Hopman}}{2006}]{Hop06c}
{Hopman} C.,  2006, astro-ph/0608460

\bibitem[\protect\citeauthoryear{{Hopman}}{{Hopman}}{2009}]{Hop09}
{Hopman} C.,  2009, ArXiv e-prints, arXiv:0906.0374

\bibitem[\protect\citeauthoryear{{Hopman}, {Freitag} \& {Larson}}{{Hopman} et~al.}{2007}]{Hop07}
{Hopman} C., {Freitag} M., {Larson} S.~L., 2007, \mnras, 378, 129

\bibitem[\protect\citeauthoryear{{Ivanov}}{{Ivanov}}{2002}]{Iva02}
{Ivanov} P.~B.,  2002, \mnras, 336, 373

\bibitem[{{Keshet} {et~al.}(2009){Keshet}, {Hopman}, \& {Alexander}}]{Kes09}
{Keshet}, U., {Hopman}, C., {Alexander}, T. 2009, ArXiv e-prints, arXiv:0901.4343

\bibitem[\protect\citeauthoryear{{Larson}}{{Larson}}{2001}]{Lar01}
{Larson} S.~L., Online Sensitivity Curve
Generator, 2001, based on
Larson, S.~L., Hellings, R.~W.\ \& Hiscock, W.~A., 2002, Phys.; located at
\texttt{http://www.srl.caltech.edu/$\sim$shane/sensitivity/} \
Rev.\ D, 66, 062001.

\bibitem[\protect\citeauthoryear{{Lightman} \& {Shapiro}}{{Lightman} \&
 {Shapiro}}{1977}]{Lig77}
{Lightman} A.~P.,  {Shapiro} S.~L.,  1977, \apj, 211, 244
\bibitem[\protect\citeauthoryear{{Maness}, {Martins}, {Trippe}, {Genzel}, {Graham}, {Sheehy}, {Salaris}, {Gillessen}, {Alexander}, {Paumard}, {Ott}, {Abuter}, \& {Eisenhauer}}{{Maness} et~al.}{2007}]{Man07}{Maness} Hl, et~al., 2007, 669, 1024

\bibitem[\protect\citeauthoryear{{Merritt} \& {Ferrarese}}{{Merritt} \& {Ferrarese}}{2001}]{Mer01}
{Merritt} D., {Ferrarese} L., 2001, \mnras, 320, L30

\bibitem[\protect\citeauthoryear{{Mouawad}, {Eckart}, {Pfalzner},
  {Sch{\"o}del}, {Moultaka} \& {Spurzem}}{{Mouawad} et~al.}{2005}]{Mou05}
{Mouawad} N.,  {Eckart} A.,  {Pfalzner} S.,  {Sch{\"o}del} R.,  {Moultaka} J.,
    {Spurzem} R.,  2005, Astron. Nachr., 326, 83

\bibitem[\protect\citeauthoryear{{Nayakshin} \& {Sunyaev}}{{Nayakshin} \& {Sunyaev}}{2005}]{Nay05}
{Nayakshin} S., {Sunyaev} R., 2005, \mnras, 364, L23

\bibitem[\protect\citeauthoryear{{O'Leary}, {Kocsis} \& {Loeb}}{{O'Leary}
  et~al.}{2008}]{OLe08}
{O'Leary} R.~M.,  {Kocsis} B.,    {Loeb} A.,  2008, ArXiv e-prints, 807

\bibitem[\protect\citeauthoryear{{Peebles}}{{Peebles}}{1972}]{Pee72} {Peebles} P.~J.~E., 1972, \apj, 178, 371 

\bibitem[\protect\citeauthoryear{{Peters} \& {Mathews}}{{Peters} \& {Mathews}}{1963}]{Pet63}
{Peters} P.~C., {Mathews} J., 1963, Phys. Rev., 131, 435

\bibitem[\protect\citeauthoryear{{Quinlan}, {Hernquist} \& {Sigurdsson}}{{Quinlan} et~al.}{1995}]{Qui95}
{Quinlan} G.~D., {Hernquist} L., {Sigurdsson} S., 1995, \apj, 440, 554

\bibitem[\protect\citeauthoryear{{Rauch} \& {Tremaine}}{{Rauch} \&
 {Tremaine}}{1996}]{Rau96}
{Rauch} K.~P.,  {Tremaine} S.,  1996, New Astron., 1, 149

\bibitem[\protect\citeauthoryear{{Rauch} \& {Ingalls}}{{Rauch} \&
 {Ingalls}}{1998}]{Rau98}
{Rauch} K.~P.,  {Ingalls} B.,  1998, \mnras, 299, 1231

\bibitem[\protect\citeauthoryear{{Rubbo}, {Holley-Bockelmann} \&
 {Finn}}{{Rubbo} et~al.}{2006}]{Rub06}
{Rubbo} L.~J.,  {Holley-Bockelmann} K.,    {Finn} L.~S.,  2006, \apjl, 649, L25


\bibitem[{{Shapiro} \& {Marchant}(1978)}]{Sha78}
{Shapiro}, S.~L., \& {Marchant}, A.~B. 1978, \apj, 225, 603


\bibitem[\protect\citeauthoryear{{Sigurdsson} \& {Rees}}{{Sigurdsson} \&
 {Rees}}{1997}]{Sig97}
{Sigurdsson} S.,  {Rees} M.~J.,  1997, \mnras, 284, 318

\bibitem[\protect\citeauthoryear{{Spitzer} \& {Saslaw}}{{Spitzer} \& {Saslaw}}{1966}]{Spi66}
{Spitzer} L.~J., {Saslaw} W.~C., 1966, \apj, 143, 400

\bibitem[{{Spitzer}(1987)}]{Spi87}
{Spitzer}, L. 1987, {Dynamical evolution of globular clusters}, ed.
  L.~{Spitzer}


\bibitem[\protect\citeauthoryear{{Tremaine},  S. and others} {{Tremaine} et~al.}{2002}]{Tre02}{Tremaine} S., et~al., 2002, \apj, 574, 704

\bibitem[\protect\citeauthoryear{{Young}}{{Young}}{1980}]{You80}
{Young} P., 1980, \apj, 242, 1232

\bibitem[\protect\citeauthoryear{{Yunes}, {Sopuerta}, {Rubbo}, \& {Holley-Bockelmann}}{{Yunes} et~al.}{2008}]{Yun08}
{Yunes} N., {Sopuerta} C.~F.,  {Rubbo} L.~J.,  {Holley-Bockelmann} K.,  2008, \apj, 675, 604
\end{thebibliography}

%\bsp
\label{lastpage}
\end{document}